%% file: VB_eprint_dpf2015.tex
\documentclass[12pt]{article}
\usepackage{graphicx}

\newcommand\pubnumber{DPF2015-224}
\newcommand\pubdate{\today}

\def\pnnl{Pacific Northwest National Laboratory\\
902 Battelle Boulevard, 99352 - Richland, WA, USA,\\
for the Belle II Collaboration.}

\def\Title#1{\begin{center} {\Large #1 } \end{center}}
\def\Author#1{\begin{center}{ \sc #1} \end{center}}
\def\Address#1{\begin{center}{ \it #1} \end{center}}

\newcommand\pubblock{\rightline{\begin{tabular}{l} \pubnumber\\
         \pubdate  \end{tabular}}}
\newenvironment{Abstract}{\begin{quotation}  }{\end{quotation}}
\newenvironment{Presented}{\begin{quotation} \begin{center} 
             PRESENTED AT\end{center}\bigskip 
      \begin{center}\begin{large}}{\end{large}\end{center} \end{quotation}}

\input econfmacros.tex
\begin{document}
\begin{titlepage}
\pubblock

\vfill
\Title{GRID Computing at Belle II }
\vfill
\Author{ Vikas Bansal}
\Address{\pnnl}
\vfill
\begin{Abstract}
The Belle II experiment at the SuperKEKB collider in Tsukuba, Japan, will start physics data taking 
in 2018 and will accumulate 50 ab$^{-1}$ of e$^{+}$e$^{-}$ collision data, about 50 times larger than the data 
set of the earlier Belle experiment. The computing requirements of Belle II are comparable 
to those of a run I high-p$_T$ LHC experiment. Computing will make full use of such grids 
in North America, Asia, Europe, and Australia, and high speed networking. 
Results of an initial MC simulation campaign with 3 ab$^{-1}$ equivalent luminosity will be described.
\end{Abstract}
\vfill
\begin{Presented}
DPF 2015\\
The Meeting of the American Physical Society\\
Division of Particles and Fields\\
Ann Arbor, Michigan, August 4--8, 2015\\
\end{Presented}
\vfill
\end{titlepage}
\def\thefootnote{\fnsymbol{footnote}}
\setcounter{footnote}{0}

\section{Introduction}

The Belle II experiment is the successor of the Belle 
experiment~\cite{BelleExp} at the KEK laboratory in Tsukuba, Japan. 
The Belle experiment measured charge-parity (CP) violation in the $B^{0}$ system 
predicted by the theory of 
Kobayashi and Maskawa~\cite{KobMas}. 
The successful confirmation of the prediction led to the Nobel Prize to both theorists.

The standard model of particle physics is still incomplete in describing
 nature and any indication of new kind of CP violation will bring us
closer to the well-known puzzle of matter-antimatter asymmetry in the
present universe.  
The KEKB accelerator delivered a total integrated luminosity of 1 ab$^{-1}$ to 
Belle experiment.
We aim to collect a data sample of 50 ab$^{-1}$ with the upgraded
KEKB accelerator, SuperKEKB, until the year 2025, primarily
to extend searches
for new kinds of CP violations.

Precision flavor physics measurements to be performed by Belle II are complementary to
the direct search for new particles at the LHC.
If new physics is found at the LHC, flavor physics
measurements are essential to identify the kind of new physics.

\section{Belle II Computing Model}

The Belle II collaboration was officially founded in December 2008. 
Today, it has more than 600
members from over 98 institutes in 23 different countries.
With collaborators located in North America, Asia, Europe, and 
Australia it is distributed around the world.

Beam collisions are expected to start in 2017. 
A data sample of about 50 times the size collected by the
Belle experiment is expected to be recorded by Belle II in 10 years from now.
Its data rate is predicted to be on par 
with the LHC~\cite{LHC}. 
Table~\ref{tab:ResourcesYear} shows an estimation
of the Belle II computing resources through 2018. 
These resources are distributed across the collaboration according
 to Belle II distributed computing model. 

%%%%%%%%%%%%%%%%%%%%%%%%%%%%%%%%%%%%%%%%%%%%%%%%%%%%%%%%%%%%%%%%%%%%%%%%%
%%
%%   use this format to include a LaTeX table  into your paper
%%
\begin{table}[t]
\begin{center}
\begin{tabular}{l|ccc}  
 		& 	2016 	&  	2017 &  2018 \\ \hline
  Disk [TB ]& 	4000        & 8000           & 9000      \\
  Tape [TB ]& 	1000        & 3000           & 5000      \\
  CPU [kHepSPEC06 ]& 	200        & 300           & 350      \\
\end{tabular}
\caption{Belle II Computing Resources Expectation}
\label{tab:ResourcesYear}
\end{center}
\end{table}
%%%%%%%%%%%%%%%%%%%%%%%%%%%%%%%%%%%%%%%%%%%%%%%%%%%%%%%%%%%%%%%%%%%%%%%%%%%

\begin{figure}[htb]
\centering
\includegraphics[height=4.0in]{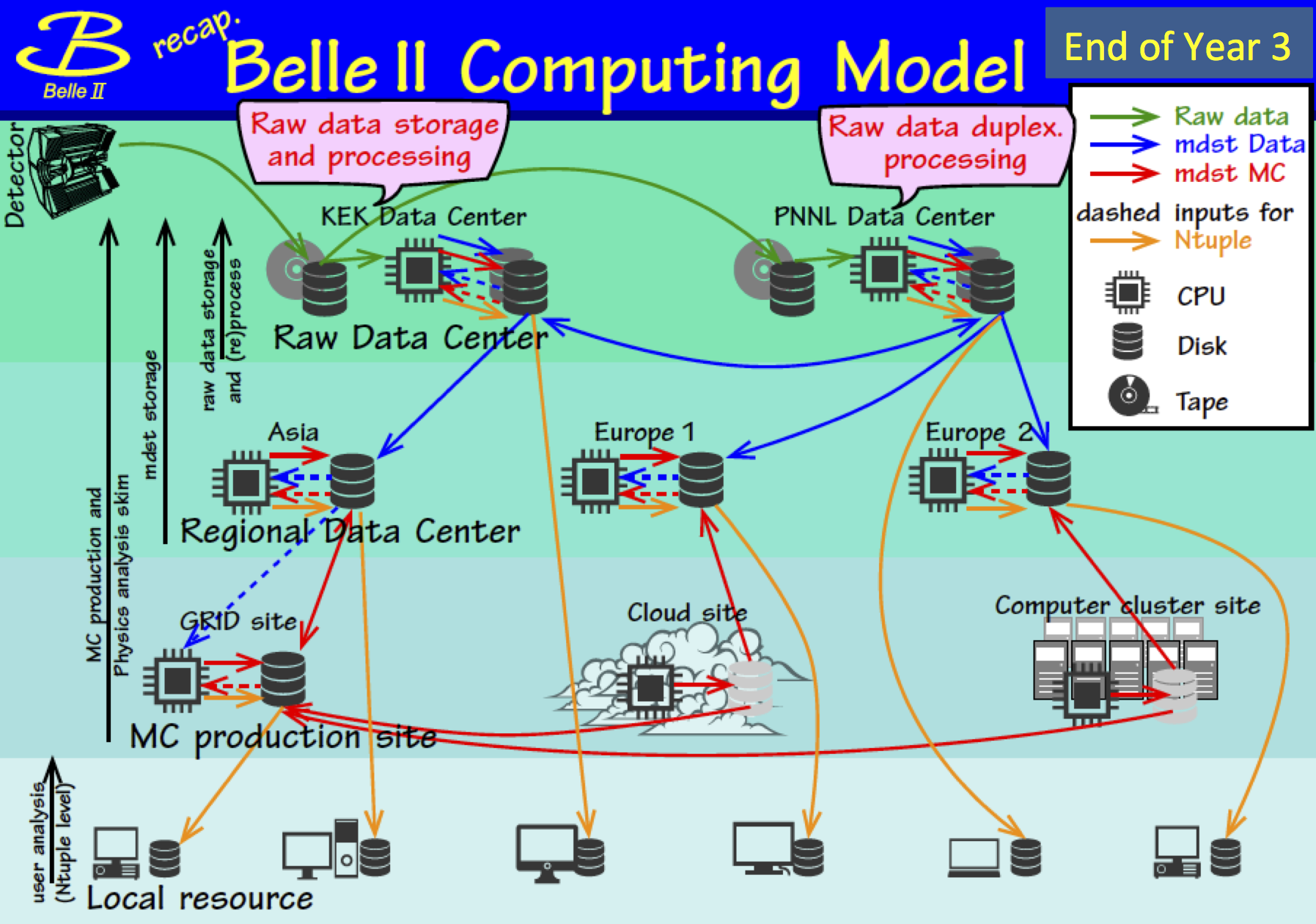}
\caption{Belle II Computing Model for the first three years of operations.}
\label{fig:BelleComputeModel}
\end{figure}

Figure~\ref{fig:BelleComputeModel} illustrates the Belle II
computing model for the first three years of operations.
PNNL along with KEK will host a compete replica of raw data
and will help to distribute processed data to Europe.

%%%%%%%%%%%%%%%%%%%%%%%%%%%%%%%%%%%%%%%%%%%%%%%%%%%%%%%%%%%%%%%%%%%%%%%%%
%%
%%   use this format to include an .pdf figure into your paper
%%
%%%%%%%%%%%%%%%%%%%%%%%%%%%%%%%%%%%%%%%%%%%%%%%%%%%%%%%%%%%%
%%%%%%%%%%%%%%%

To ensure that appropriate bandwidths are available on network routes
 between major sites in North America, Asia, Europe, and 
Australia, they are routinely 
tested for network traffic throughput rates as part of data challenges.
Virtual LAN (VLAN) setups were established between sites to perform
network data challenges both trans-atlantic and trans-pacific using 
FTS3~\cite{FTS3} service.
A site-to-site matrix was deployed to monitor and capture network information for
latency and network packet drop rates using MaDDash~\cite{MaDDash}
 and perfSONAR~\cite{perfSONAR} services.
Network information between sites is inserted in distributed database that guides
optimal data routing path between any two endpoints.

Belle II was recently included in the LHCONE~\cite{LHCONE}
 Acceptable Use Policy (AUP).
This will enable Belle II member sites to interconnect via the
well managed infrastructure of
 LHC network and hence provide maximal throughput possible.
A key component of the LHCONE is that all the sites are ``trusted'' which
alleviates the need for firewalls (as it slows down data transfer rates).
Also, bandwidths via public internet is limited and not feasible
to support high data transfer rates.

%One ubiquitous thing at Belle II, as at any HEP experiment, is to
%continuosly store and process data.
Belle II has chosen DIRAC~\cite{DIRAC} to
provide key functionality for their distributed computing model.
DIRAC will orchestrate all its sub-components that can 
as well be distributed 
at many sites to achieve Belle II compuitng operations.
It will also manage computing resources across all sites.

 Belle II collaboration 
is developing various components and extensions to DIRAC, 
including a fabrication system 
that manages compute tasks over all sites, a distributed data management system, 
web portals for monitoring purposes, 
and dedicated wrappers around the Belle II reconstruction framework.

\begin{figure}[htb]
\centering
\includegraphics[height=4.0in]{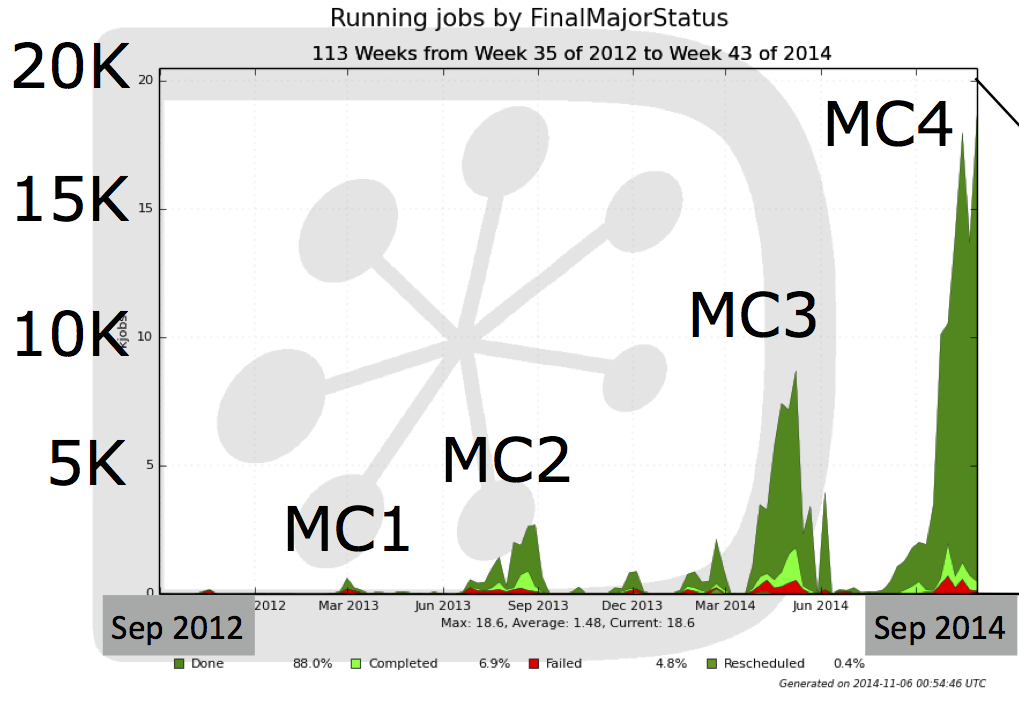}
\caption{MC Jobs at Belle II MC production campigns between 2012 and 2014. 
Y-axis shows number of production jobs and X-axis spans time.}
\label{fig:MCCampaigns_4}
\end{figure}

The simulation samples for the Belle II experiment 
have been produced in a globally distributed manner, 
in accordance with the distributed computing model.
Figure~\ref{fig:MCCampaigns_4}
illustrates the output of the DIRAC-based production system for four
MC production campaigns between 2012 and 2014.
All produced data is again distributed to storage resources
at various sites.

\section{Conclusions}

The large increase in sensitivity 
of Belle II over Belle
comes with a significantly larger data sample.
The SuperKEKB accelerator is designed to provide 50 times more data until the year 2025.
This data volume is on par with the LHC and is a challenge for the computing system.
The strategy involves PNNL to host a complete replica of 
raw data for the first three years in operation
and distribute processed data to Europe. 
To ensure optimal network bandwidths, we routinely 
perform data challenges over the Belle II network which is now
part of LHCONE AUP.
Our computing model is facilitated by DIRAC and we have successfully 
tested the prototype over MC production campaigns.
More details about the computing at Belle II can be found in Chapter 14 of the
Belle II Technical Design Report~\cite{BELLE2TDR}. 

%\Acknowledgments
%Thanks to Belle II collaboration.

\end{document}

%% file: econfmacros.tex
\def\beq{\begin{equation}}
\def\eeq#1{\label{#1}\end{equation}}
\def\eeqn{\end{equation}}

\def\beqa{\begin{eqnarray}}
\def\eeqa#1{\label{#1}\end{eqnarray}}
\def\eeqan{\end{eqnarray}}

\let\bar=\overbar

\def\Dslash{\not{\hbox{\kern-4pt $D$}}}
\def\dslash{\not{\hbox{\kern-2pt $\del$}}}

\def\msb{{\bar{\ssstyle M \kern -1pt S}}}